\documentclass[%
 reprint,
superscriptaddress,
 amsmath,amssymb,
 aps,
 prl,
]{revtex4-2}

\usepackage{graphicx}
\usepackage{dcolumn}
\usepackage{bm}
\UseRawInputEncoding

\begin{document}


\title{Pure nematic state in iron-based superconductor}

\author{Y. Kubota}
 \email{kubota@spring8.or.jp}
 \affiliation{RIKEN SPring-8 Center, 1-1-1 Kouto, Sayo, Hyogo 679-5148, Japan}
 
\author{F. Nabeshima}
\affiliation{Department of Basic Science, University of Tokyo, 3-8-1 Komaba, Meguro, Tokyo 153-8902, Japan}

\author{K. Nakayama}
\affiliation{Department of Physics, Graduate School of Science, Tohoku University, Sendai 980-8578, Japan}
\affiliation{Precursory Research for Embryonic Science and Technology, Japan Science and Technology Agency, Tokyo 102-0076, Japan}

\author{H. Ohsumi}
\affiliation{RIKEN SPring-8 Center, 1-1-1 Kouto, Sayo, Hyogo 679-5148, Japan}

\author{Yoshikazu Tanaka}
\affiliation{RIKEN SPring-8 Center, 1-1-1 Kouto, Sayo, Hyogo 679-5148, Japan}

\author{K. Tamasaku}
\affiliation{RIKEN SPring-8 Center, 1-1-1 Kouto, Sayo, Hyogo 679-5148, Japan}
\affiliation{Japan Synchrotron Radiation Research Institute (JASRI), 1-1-1 Kouto, Sayo, Hyogo 679-5198, Japan}

\author{T. Suzuki}
\affiliation{Institute for Solid State Physics, The University of Tokyo, Kashiwa, Chiba 277-8581, Japan}

\author{K. Okazaki}
\affiliation{Institute for Solid State Physics, The University of Tokyo, Kashiwa, Chiba 277-8581, Japan}

\author{T. Sato}
\affiliation{Department of Physics, Graduate School of Science, Tohoku University, Sendai 980-8578, Japan}
\affiliation{Advanced Institute for Materials Research (WPI-AIMR), Tohoku University, Sendai 980-8577, Japan}

\author{A. Maeda}
\affiliation{Department of Basic Science, University of Tokyo, 3-8-1 Komaba, Meguro, Tokyo 153-8902, Japan}

\author{M. Yabashi}
\affiliation{RIKEN SPring-8 Center, 1-1-1 Kouto, Sayo, Hyogo 679-5148, Japan}
\affiliation{Japan Synchrotron Radiation Research Institute (JASRI), 1-1-1 Kouto, Sayo, Hyogo 679-5198, Japan}




\date{\today}

\begin{abstract}
Lattice and electronic states of thin FeSe films on LaAlO$_3$ substrates are investigated in the vicinity of the nematic phase transition.
No evidence of structural phase transition is found by x-ray diffraction below $T^\ast \sim 90$~K, while results obtained from resistivity measurement and angle-resolved photoemission spectroscopy clearly show the appearance of a nematic state.
These results indicate formation of a pure nematic state in the iron-based superconductor and provide conclusive evidence that the nematic state originates from the electronic degrees of freedom.
This pure nematicity in the thin film implies difference in the electron-lattice interaction from bulk FeSe crystals.
FeSe films provide valuable playgrounds for observing the pure response of ``bare" electron systems free from the electron-lattice interaction, and should make important contribution to investigate nematicity and its relationship with superconductivity.
\end{abstract}

\maketitle


Iron-based superconductors have been extensively studied since the discovery of superconductivity in LaFeAsO$_{1-y}$F$_y$~\cite{Kamihara2008, Stewart2011}.
Iron-chalcogenide superconductors such as FeSe have also been attracting a great interest owing to the simplest crystal structure and the absence of antiferromagnetism compared with other iron-based superconductors~\cite{Hsu2008, Bohmer2018, Coldea2018, Kreisel2020, Shibauchi2020}.
Furthermore, FeSe has exotic features concerning the critical temperature ($T_{\mathrm{c}}$).
Although $T_{\mathrm{c}}$ is only $\sim 8$~K at ambient pressure~\cite{Hsu2008}, $T_{\mathrm{c}}$ increases to $40$--$50$~K by physical pressure~\cite{Mizuguchi2008, Medvedev2009, Miyoshi2014, Kothapalli2016, Sun2016}, chemical manipulations~\cite{Guo2010, Miyata2015, Wen2016, Ying2012, Burrard2013, Hosono2014}, and electric field~\cite{Shiogai2016, Hanzawa2016, Lei2016}.
Monolayer FeSe films exhibit $T_{\mathrm{c}}$ above $65$~K~\cite{Wang2012, He2013}.
To understand the origin of Cooper pair formation in iron-based superconductors, it is important to reveal the relationship between superconductivity and exotic properties owing to the electronic degrees of freedom.

\begin{figure}
\begin{center}
\includegraphics[width=8cm]{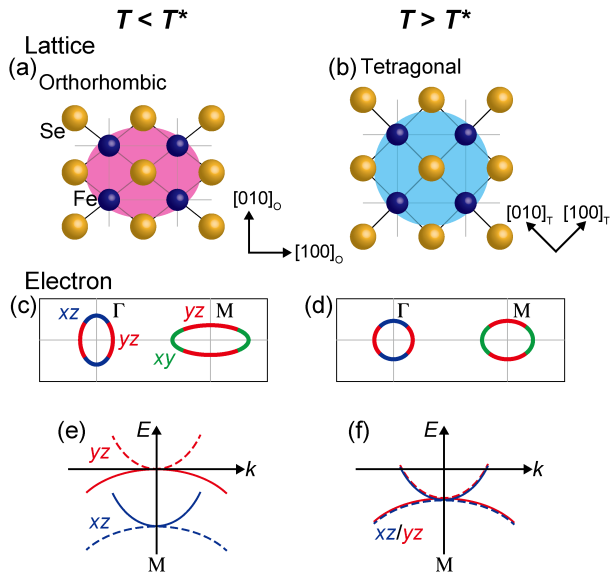}
\caption{
Comparisons of (a), (b) lattice structures, (c), (d) Fermi surfaces around the $\Gamma$ and $M$ points, and (e), (f) band diagrams around the $M$ point of FeSe between the nematic state ($T < T^\ast$) and the normal state ($T > T^\ast$).
The pink and blue circles in (a) and (b) indicate the appearance of electronic nematicity.
The blue, red, and green curves  in (c)--(f) indicate the $d_{xz}$, $d_{yz}$, and $d_{xy}$ orbital bands, respectively.
Solid and dashed curves in (e) and (f) represent the band dispersions along the $(0, 0)$--$(\pi, 0)$ and $(0, 0)$--$(0, \pi)$ directions (long and short Fe--Fe directions), respectively, of the untwinned crystal~\cite{Nakayama2014}.
}
\label{fig1} 
\end{center}
\end{figure}
Electronic nematicity, where the electronic system spontaneously breaks a rotation symmetry accompanied by a structural change and a development of antiferromagnetism, is an intriguing state shown in almost all families of iron-based superconductors~\cite{Chuang2010, Chu2010, Tanatar2010, Dusza2011, Yi2011, Kasahara2012, Chu2012, Fernandes2014}.
Figure~\ref{fig1} shows the variations in the lattice and band structures of FeSe at the nematic transition.
The lattice undergoes a tetragonal-to-orthorhombic structural transition at $T^\ast \sim 90$~K~\cite{Hsu2008, Margadonna2008, McQueenPRL2009, Bohmer2013}.
The shape of the Fermi surface and the band diagram change across $T^\ast$ because of lifting of the $d_{xz}$ and $d_{yz}$ orbital degeneracy~\cite{Nakayama2014, Shimojima2014, Tan2016, Phan2017, Zhang2015, Watson2015, Suzuki2015, Yi2019}.
Around the $M$ point, the $d_{xz}$ band shifts downward along the $(0, 0)$--$(0, \pi)$ direction of the untwinned crystal, and the $d_{yz}$ band shifts upward along the $(0, 0)$--$(\pi, 0)$ direction; this leads to the breaking of the four-fold rotational symmetry.
For the twinned FeSe sample, these two bands can be observed simultaneously, as shown in Figs.~\ref{fig1}(e) and \ref{fig1}(f).
Because the nematic phase is adjacent to the superconducting phase in the phase diagram, nematicity is considered to be closely related to superconductivity~\cite{Fernandes2014, Bohmer2018, Coldea2018, Kreisel2020, Shibauchi2020}.
Investigation of the origin and physical properties of nematicity will therefore lead to clarification of unsettled superconducting mechanisms.
Which degrees of freedom are the origin of nematicity is an important issue.

In this Letter, we have investigated the lattice and electronic properties of thin FeSe films on LaAlO$_3$ (LAO) substrates in the vicinity of the nematic phase transition with x-ray diffraction (XRD), resistivity measurement, and angle-resolved photoemission spectroscopy (ARPES).
FeSe is an ideal platform for studying the origin of nematicity because it undergoes the structural transition without any magnetic order~\cite{Dai2015}.
Even though there is a growing consensus that the electronic degrees of freedom contribute to nematicity, to what extent and how the lattice degrees of freedom contribute are far from a complete understanding due to several conflicting reports.
For instance, a minor role of lattice degrees of freedom was suggested because the energy splitting of the electronic system in the nematic state is larger than that estimated from the lattice variation~\cite{Fernandes2014}, whereas an intimate role is anticipated from a strongly intertwined nature of the nematic state with the structural transition in almost all iron-based superconductors.
Also, a recent study suggested that the contribution from the structural transition to nematicity cannot be negligible~\cite{Yang2022}.
Therefore, the interplay between nematicity and the lattice degrees of freedom is controversial, partly because of the difficulty in completely decoupling the electron and lattice degrees of freedom.
Our observation of a pure electronic nematic state unaccompanied by the structural phase transition provides conclusive evidence that the nematic state originates from the electronic degrees of freedom, i.e. the lattice degrees of freedom are not essential to realize nematicity.

\begin{figure}
\begin{center}
\includegraphics[width=8cm]{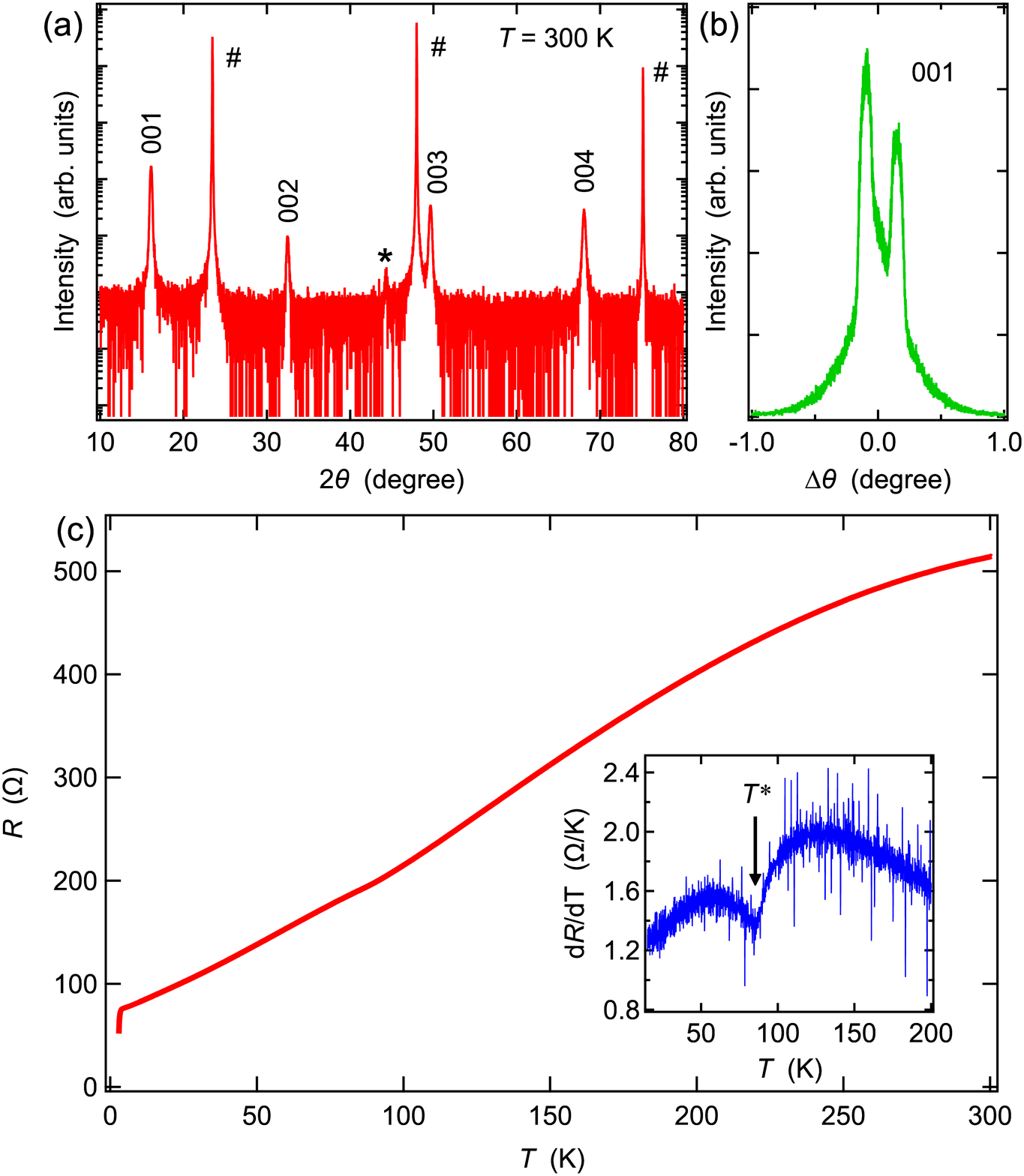}
\caption{
Characterization of FeSe film on LAO.
(a) XRD pattern obtained at room temperature.
The number and asterisk represent peaks from the LAO substrate and the stainless-steel sample holder, respectively.
(b) Rocking curve of the $0 0 1$ reflection of the FeSe film.
(c) Temperature dependence of the resistance $R$ of the FeSe film.
The inset shows the temperature dependence of d$R /$d$T$ of the FeSe film.
The arrow shows the nematic transition temperature $T^\ast$.
}
\label{fig2}
\end{center}
\end{figure}
The FeSe thin films in this study were grown on LAO $(0 0 1)$ substrates using a pulsed laser deposition method with a KrF laser ($\lambda = 248$~nm)~\cite{Imai2010, Imai2010_2}.
The growth temperature and growth rate were $500^\circ$C and $\sim 1$~nm/min, respectively.
The thickness of the film for XRD measurements was approximately $50$~nm.
Much thicker films ($> 300$~nm) were prepared for ARPES measurements to enable easy cleavage of the films.
Figure~\ref{fig2}(a) shows the XRD $\theta$--$2\theta$ scan at room temperature of an FeSe film on LAO; the scan shows the $c$-axis orientation of the film.
The $c$-axis lattice constant is $5.50$~$\mathrm{\AA}$, which is slightly shorter than those of bulk samples ($5.521$--$5.525$~$\mathrm{\AA}$~\cite{McQueenPRB2009, Kasahara2014}).
The shorter $c$-axis indicates an in-plane tensile strain in the film~\cite{Nabeshima2018}.
Figure~\ref{fig2}(b) shows the rocking curve of the $0 0 1$ reflection of the film.
The two-peak structure is due to the twin of the rhombohedral LAO substrate.

ARPES was performed with Scienta-Omicron-DA30 and MBS-A1 electron analyzers at BL28A in Photon Factory~\cite{Kitamura2022} and BL7U in UVSOR, respectively.
Photoelectrons were excited by using circularly polarized $56$-eV photons and linearly polarized $21$-eV photons.
The energy and angular resolutions were set to be $12$--$30$ meV and $0.3$~degrees, respectively.
A clean surface for ARPES measurements was obtained by cleaving FeSe films in situ in an ultrahigh vacuum, i.e., better than $1 \times 10^{-10}$~Torr.
The thickness of the films after the cleavage was estimated to be $\sim 70$~nm from transmission electron microscopy measurements (see Supplemental Material).
The Fermi level ($E_\mathrm{F}$) of the films was referenced to that of a gold film electrically contacted with the sample holder.

Low-temperature XRD was performed at BL19LXU of SPring-8~\cite{Yabashi2001}.
The photon energy of the x-ray beam was $10$~keV monochromatized by a pair of cryogenically cooled Si $(1 1 1)$ crystals.
The energy resolution was $\sim 1 \times 10^{-4}$.
The x-ray intensity of the $3 3 6$ Bragg reflection in the tetragonal structure was detected with a NaI scintillation counter and an x-ray CMOS image sensor, SOPHIAS~\cite{Hatsui2013}.
This high-index reflection was chosen to enable detection of changes in the lattice structure with high precision.
The distance between the sample and SOPHIAS was $\sim 2$~m, which resulted in an angle resolution of $8.6 \times 10^{-4}$~degrees.
The sample was cooled with a cryostat to a minimum of $6$~K.

Figure~\ref{fig2}(c) shows the temperature dependence of the resistance $R$ of an FeSe film on LAO, which was grown at the same time as the sample for the XRD measurement at low temperatures.
Because of the tensile strain, the superconducting transition temperature $T_c^{\mathrm{onset}}$ (3.2 K) was lower than those of bulk samples~\cite{Nabeshima2018, Nakajima2021}.
A resistive anomaly, which is clearly indicated by a dip in the temperature derivative of the resistance [inset in Fig.~\ref{fig2}(c)], suggests the nematic transition at $T^\ast \sim 90$~K~\cite{Shimojima2014}.

\begin{figure*}
\begin{center}
\includegraphics[width=18cm]{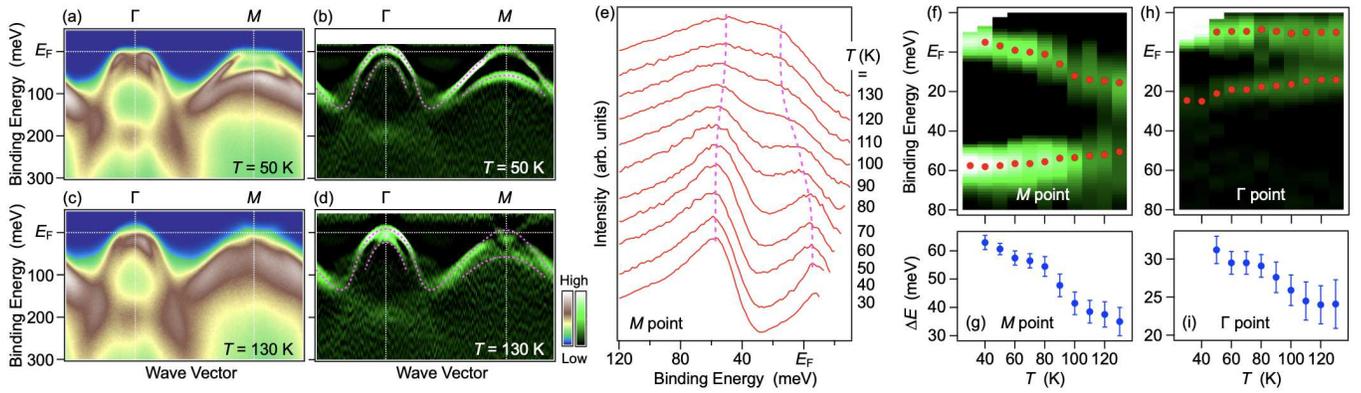}
\caption{
(a) ARPES intensity as a function of binding energy and wave vector along the $\Gamma$--$M$ cut at $T = 50$~K with $56$-eV photons.
(b) Second-derivative intensity of (a), after division by the Fermi-Dirac distribution function convoluted with the instrumental resolution function.
Magenta dashed curves are a guide for the eyes to trace the band dispersions near $E_\mathrm{F}$.
(c), (d) Same as (a) and (b), respectively, but at $T = 130$~K.
Magenta dashed curves in (d) are identical to those in (b).
(e), (f) Temperature dependence of ARPES spectrum at the $M$ point and the corresponding second-derivative intensity plot, respectively.
Magenta dashed curves in (e) are a guide for the eye to trace the energy position of $d_{xz} / d_{yz}$-derived bands, extracted from the peak position in (f) (red circles).
(g) Magnitude of the energy difference between the $d_{xz}$- and $d_{yz}$-derived bands at the $M$ point plotted as a function of temperature.
(h), (i) Same as (f) and (g), respectively, but for the $\Gamma$ point.
}
\label{fig3}
\end{center}
\end{figure*}
To corroborate the emergence of nematicity in the FeSe films on LAO, we have investigated the evolution of the electronic structure across $T^\ast$ by using ARPES.
Figure~\ref{fig3}(a) shows a representative ARPES intensity plot at low temperatures, i.e., below $T^\ast$ ($T = 50$~K), along the $\Gamma$--$M$ cut of the Brillouin zone.
The corresponding second-derivative intensity is plotted in Fig.~\ref{fig3}(b).
Around the $M$ point, there are two holelike bands with the top at binding energies $E_\mathrm{B}$s of $\sim E_\mathrm{F}$ and $\sim 60$~meV (see magenta dashed curves).
On the basis of previous studies of FeSe bulk crystals and thin films, these bands are attributed to the Fe $3d_{xz} / d_{yz}$ orbitals~\cite{Nakayama2014, Shimojima2014, Tan2016, Phan2017}.
A large energy separation ($\Delta E \sim 60$~meV) between the $d_{xz} / d_{yz}$ orbitals at the $M$ point is a hallmark of the nematic state [Fig.~\ref{fig1}(e)]~\cite{Nakayama2014, Shimojima2014, Tan2016, Phan2017, Zhang2015, Watson2015, Suzuki2015, Yi2019} and therefore supports the formation of nematicity at low temperatures in the FeSe films on LAO (note that the observation of both $d_{xz}$- and $d_{yz}$-derived holelike bands suggests that the two nematic domains perpendicular to each other are simultaneously probed in the present experiment).
The ARPES intensity plots at $T = 130$~K in Figs.~\ref{fig3}(c) and \ref{fig3}(d) show that $\Delta E$ at the $M$ point, i.e., nematicity, is suppressed at high temperatures.
We examined this point in detail by plotting the temperature dependence of the ARPES spectrum at the $M$ point and its second-derivative intensity, as shown in Figs.~\ref{fig3}(e) and \ref{fig3}(f), respectively.
When the temperature was increased, the band at higher $E_{\mathrm{B}}$ shifted toward $E_{\mathrm{F}}$, whereas the band at lower $E_{\mathrm{B}}$ shifted away from $E_{\mathrm{F}}$, which led to a decrease in $\Delta E$.
Note that a finite $\Delta E$ at high temperatures likely reflects a gap opening associated with the spin-orbit coupling (SOC)~\cite{Borisenko2016} and/or a contribution from the $d_{xy}$ orbital~\cite{Yi2019, Watson2016, Rhodes2021}.
The temperature dependence of $\Delta E$ shown in Fig.~\ref{fig3}(g) indicates that nematicity in the FeSe films on LAO is gradually suppressed in the temperature range $80$--$100$~K.
Around the $\Gamma$ point, there are two holelike bands that also have $d_{xz} / d_{yz}$ orbital characteristics with $\Delta E \sim 30$~meV at $T = 50$~K [Fig.~\ref{fig3}(b)].
Similar to the observation around the $M$ point, $\Delta E$ at the $\Gamma$ point shows a finite decrease upon raising the temperature to $130$~K [Fig.~\ref{fig3}(d); see an upward energy shift of the lower branch compared to the band dispersion at $50$~K (magenta dashed curve)], indicating that the $d_{xz} / d_{yz}$ orbitals at the $\Gamma$ point are also influenced by nematicity, in agreement with the previous study~\cite{Zhang2016}.
The detailed temperature-dependent study at the $\Gamma$ point [Figs.~\ref{fig3}(h) and \ref{fig3}(i)] confirms a gradual suppression of nematicity in the temperature range of $80$--$100$~K.
All these results strongly support the emergence of nematicity in the FeSe films on LAO, as estimated from the electrical resistivity measurement [Fig.~\ref{fig2}(c)].

\begin{figure}
\begin{center}
\includegraphics[width=9cm]{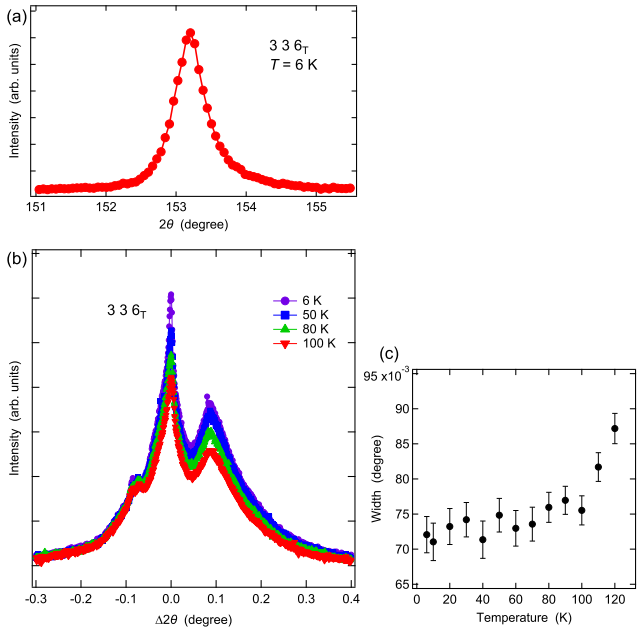}
\caption{
(a) Result of $\theta$--$2\theta$ scan at the $3 3 6_{\mathrm{T}}$ Bragg reflection obtained at $6$~K.
(b) Temperature dependence of the XRD profile at the $3 3 6_{\mathrm{T}}$ Bragg reflection obtained with SOPHIAS.
The $\theta$ angle was fixed at the maximum diffraction signal at each temperature.
The horizontal axis shows the value relative to the main peak angle.
(c) Temperature dependence of the width of the main peak obtained by the fitting with the Lorentz function for the XRD profiles.
The vertical bars represent error bars from the fitting.
}
\label{fig4}
\end{center}
\end{figure}
Figure~\ref{fig4} shows the results of the XRD measurement at the $3 3 6$ Bragg reflection in the tetragonal structure.
Hereafter, we describe this reflection as $3 3 6_{\mathrm{T}}$, while the reflections in the orthorhombic structure use the subscript of O.
We measured the temperature dependence of diffraction from $6$~K to $120$~K.
In bulk samples, the $3 3 6_{\mathrm{T}}$ peak has been known to split into a doublet, i.e., $6 0 6_{\mathrm{O}}$ and $0 6 6_{\mathrm{O}}$, because of the structural transition below $T^\ast$~\cite{Hsu2008, Margadonna2008, McQueenPRL2009, Bohmer2013}.
The split angle at $T = 6$~K is evaluated to be $\sim 1.3$~degrees in $2 \theta$ for wavelength $\lambda = 0.124$~nm by using previously reported lattice constants for the bulk sample~\cite{Margadonna2008}.
However, we did not observe splitting in a $\theta$--$2\theta$ scan detected with the scintillation counter, as shown in Fig~\ref{fig4}(a).
For more precise investigation, we also measured the $3 3 6_{\mathrm{T}}$ peak with SOPHIAS placed further from the sample than the scintillation counter had been. 
Figure~\ref{fig4}(b) shows the temperature dependence of the XRD profile at the $3 3 6_{\mathrm{T}}$ with SOPHIAS.
Although the detector has a high-angle resolution of $8.6 \times 10^{-4}$~degrees, no splitting was observed.
The multiple peaks that appear regardless of temperature are due to the twin of the rhombohedral LAO substrate, as shown in Fig.~\ref{fig2}(b).
Figure~\ref{fig4}(c) shows the temperature dependence of the main peak width obtained by the fitting with the Lorentz function for the XRD profiles.
Rather than splitting, the peak width decreased with decreasing temperature.
If the peak splits, the intensity of one of the split peaks should be smaller than that of the original peak.
However, as shown in Fig.~\ref{fig4}(b), the peak intensity increased at lower temperatures (see also Supplemental Material).
There is therefore no possibility that the splitting angle is too large to be observed within the detection area.
Note that it has been confirmed that the baselines of all temperatures match in the $\Delta 2 \theta$ region far from the peak position (see Supplemental Material).
We also performed $\theta$ scans at each temperature and confirmed that there was no splitting.
From these results, we can conclude that no structural phase transition occurs in the FeSe thin film on LAO.
It is natural to consider that the structural change is suppressed by strain from the substrate~\cite{Wang2009}.

Our resistivity measurement and ARPES results are similar to those for bulk FeSe crystals, therefore there is no doubt that the nematic transition occurs at $T^\ast$ in the FeSe thin film on LAO.
However, the low-temperature XRD measurement indicates that no structural phase transition occurred.
Purely electronic nematicity is therefore realized in the FeSe thin film.
This is evidence that the nematic phase originates only from the electronic degrees of freedom.
The effective electron-lattice interaction $\lambda_{\mathrm{eff}}$ is given by $\lambda_{\mathrm{eff}} \sim \lambda^2 / \omega_{\mathrm{ph}}$, where $\lambda$ and $\omega_{\mathrm{ph}}$ are the bare electron-lattice interaction and the phonon frequency, respectively.
Assuming that the lattice deformation is suppressed by the substrate in the thin film and that $\omega_{\mathrm{ph}}$ is higher than that for bulk crystals, $\lambda_{\mathrm{eff}}$ becomes smaller, which provides a qualitative understanding of the pure nematic transition.
The pure nematic state in FeSe films provides a valuable playground in which the pure response of ``bare" electron systems free from the electron-lattice interaction can be observed.
Shimojima {\it et al.}~\cite{Shimojima2019} performed time-resolved experiments to investigate the nature of the nematic state decoupled from the lattice state by using the difference between their timescales after optical excitation.
In contrast, this study shows that the FeSe films achieve this decoupling even at the equilibrium state.
The rapid increase in $T_{\mathrm{c}}$ when the nematic transition is suppressed by changing the chemical compositions of film samples may be related to such bare nature of electrons~\cite{Imai2015, Imai2017}.
The achievement of higher $T_{\mathrm{c}}$ at ambient pressure in film samples than in bulk crystals may also suggest that phonons do not play a key role in realizing $10$--$40$~K class superconductivity in iron chalcogenides.

In summary, we discovered the pure electronic nematicity not accompanied by the lattice structural transition in the FeSe thin film on LAO.
This is evidence that the origin of nematicity is the electronic degrees of freedom.
Compared to bulk crystals, FeSe thin film has advantages in its preparation, such as easy control of composition in elemental substitutions and ease of growing single crystals with large areas~\cite{Imai2015, Imai2017, NabeshimaJPSJ2018}.
With our discovery of the pure nematic state, FeSe thin films provide an important platform for investigating the nematic state and nematic fluctuation, and their relationship with superconductivity.
The bare electron systems in FeSe film would help resolve the issue that which electronic degrees of freedom, charge/orbital or spin, drives the nematicity~\cite{Fernandes2014}.
Because the nematic phase commonly appears near the superconducting phase of many iron-based superconductors, a deeper understanding of the relationship between nematicity and superconductivity, which can be achieved by the study of the pure nematic state in FeSe/LAO, will provide an important basis for studying iron-based superconductors to clarify the mechanism of superconductivity, to increase $T_{\mathrm{c}}$, and even to realize interfacial superconductivity~\cite{Ginzburg1964, Allender1973}.

We thank Dr. Miho Kitamura, Dr. Koji Horiba, Dr. Hiroshi Kumigashira, Dr. Shin-ichiro Ideta, and Dr. Kiyohisa Tanaka for their assistance with ARPES measurements.
We also thank Ataru Ichinose at Central Research Institute of Electric Power Industry for performing the cross-sectional TEM experiment.
The XRD measurement was performed at BL19LXU of SPring-8 with the
approval of RIKEN (Proposal Nos.~20180094, 20190042, and 20200056).
This work was supported by JST-PRESTO (No. JPMJPR18L7), JST-CREST (No. JPMJCR18T1), Grant-in-Aid for Scientific Research (JSPS KAKENHI Grant Number: JP20H01847), KEK-PF (Proposal number: 2021S2-001), and UVSOR (Proposal number: 21-679).


\end{document}